\documentclass[12pt]{iopart}
\usepackage{iopams}
\usepackage{bm}
\usepackage{amsfonts}
\usepackage{graphicx}
\begin{document}

\title{Pair creation by time-dependent electric fields: Analytic solutions}
\author{Iwo Bialynicki-Birula$^{1}$, \L{}ukasz Rudnicki$^{1}$\\
 and Albert Wienczek$^{1,2}$}
\address{$^{1}$Center for Theoretical Physics, Polish Academy of Sciences\\
Al. Lotnik\'ow 32/46, 02-668 Warsaw, Poland}
\address{$^{2}$Faculty of Physics, University of Warsaw, Warsaw, Poland}

\ead{birula@cft.edu.pl}
\begin{abstract}
Exact analytical solutions are presented for the time evolution of the density of pairs produced in the QED vacuum by a uniform electric field that is adiabatically switched on starting at $-\infty$. Pair production is described by the Dirac-Heisenberg-Wigner function introduced before [Phys. Rev. D {\bf 44}, 1825 (1991)]. The explicit solution is obtained by an extension of the method of the spinorial decomposition to deal with a time-varying electric field. The main result of this work is that the pair density is an analytic function of the field strength; it can be expanded into a convergent power series. Therefore, the essential singularity present in the Schwinger formula is to be attributed to the infinitely long duration of the process of pair creation by a time-independent field.
\end{abstract}
\maketitle

\section{Introduction}

The process of pair creation by the time independent and uniform electric field as described in our previous publication \cite{bbr} suffers from a serious drawback. The instantaneous switching on of the field leads to unphysical transient effects clearly seen in Fig.~1 of \cite{bbr}. In the present paper we solve analytically the problem of pair production by a uniform electric field that is {\em switched on adiabatically}.

To explain what this means in physical terms, let us consider a large (boundary conditions are disregarded) capacitor charged so that the electric field is $\bm E$. We shall compare two cases. In the first case, the electric field is primordial, it existed always. In the second case, the capacitor has been continuously charged until at $t=0$ the field reached the desired value $\bm E$. The creation of pairs in the first case, i.e. for the time-independent electric field, was calculated by Schwinger \cite{s} whose analytical formula has a characteristic essential singularity as a function of the field strength $\mathcal E$.

Consider now the observer who at time $t=0$ can measure the strength of the electric field and the number of pairs but does not have information about the past history. This observer can plot the number of pairs as a function of the field strength and find out whether this dependence shows the singular behavior characteristic for the Schwinger solution. We show with the help of exact analytic solutions that the singularity will not be found if the field has been turned on adiabatically. In this case the Schwinger singularity is not present. The formulas that describe the total pair production for an adiabatically switched field are analytic in the field strength---they can be expanded in the field strength into a convergent power series---for every finite value of the parameter $b$, that sets the time scale of the charging process. The Schwinger singularity in the field strength appears now as a singularity in $b$. Thus, the essential singularity found by Schwinger is to be attributed entirely to rather unphysical assumptions that the field has infinite duration or that it has been suddenly switched on.

The production of pairs by an adiabatic perturbation may seem impossible since there is an energy gap of $2mc^2$ between the vacuum state and the state with one pair. Therefore, it would seem that under an adiabatic perturbation the vacuum will not change. However, these doubts should disappear when one realizes that pair production by a constant electric field is due to tunneling. Arbitrarily weak electric field can produce an arbitrarily large difference in potential energy for a sufficiently large separation. When this difference in potential energy exceeds the value $2mc^2$, pairs are created.

The problem of pair production will be formulated in terms of a field-theoretic generalization of the standard, nonrelativistic, Wigner function. This generalization was introduced in \cite{bgr} and called the Dirac-Heisenberg-Wigner (DHW) function. This formulation was further developed in \cite{sbr}-\cite{hag1}. It allows for a description of pair creation as a time evolution of the vacuum state. The advantage of using the DHW function is its applicability to the expectation values of field operators and not only to the one-particle wave functions that are solutions of the Dirac equation. However, the analytic solutions of the one-particle Dirac equation are still very useful. The method of spinorial decomposition introduced in \cite{bbr} enables one to construct from these solutions the components of the DHW function.  We shall describe two exactly soluble examples.

In the first example the field is switched on at a constant rate $b$---the time dependence is governed by a simple exponential. In the second example the field is switched on at a decreasing rate---the time dependence is governed by the hyperbolic secant squared. In both cases the signature of the emerging Schwinger singularity is the shrinking of the radius of convergence to zero when $b$ tends to zero. We also show that in the first case the power expansion of the exact solution coincides with the straightforward perturbation theory.

The evolution equation for the DHW function has many solutions, most of them do not represent physical states. The DHW function shares this property with the standard Wigner function in nonrelativistic quantum mechanics---not every function over the phase-space is a Wigner function. For example, the Wigner function cannot be concentrated too much in position and momentum since this would contradict the uncertainty relations. There is, however, a class of problems for which one can be sure that the Wigner function represents a physically admissible state. These are the problems of time evolution with the initial value being a genuine Wigner function. The DHW function that is the solution of the evolution equation will correspond to a physical state---the one  that evolved from the initial state according to the Schr\"odinger equation. This is precisely the method that is used in this work. To describe the time evolution under the influence of adiabatically switched on fields we shall always start with the easily computable exact DHW function---the one that describes the free vacuum state.

In nonrelativistic quantum mechanics the Wigner function is defined as a Fourier integral of a product of wave functions and in many cases this integral can be explicitly evaluated. In contrast, the calculation of the DHW function in quantum field theory requires the summation of contributions from an infinite number of states of virtual particles in the vacuum. The only case where this can be easily done is the vacuum state of the free field. For that reason in our analysis of the time evolution we shall choose the DHW function of the vacuum as the initial value.

In general, the evolution equations for the components of the DHW function, even in the approximation where all radiative corrections are neglected, form a set of 16 integro-differential equations and their analytic solution is a hopeless task. However, in the case of a spatially uniform electromagnetic field the evolution equations become tractable since they reduce to linear partial differential equations. These equations allow for the application of the method of characteristics leading to ordinary differential equations. In turn, these ordinary differential equations in certain cases can be solved analytically by spinorial decomposition introduced in Ref. ~\cite{bbr}.

In the present paper we construct a complete set of solutions of the evolution equations for the DHW function describing the evolution of the vacuum in the presence of a uniform but time-dependent electric field. With the use of these solutions we find the evolution when the initial state is the free-field vacuum. Our exact solutions clarify the mechanism responsible for pair production and may explain some details of the vacuum decay studied recently in the semiclassical approximation \cite{dd1}-\cite{dd3}.

In Sec.~\ref{secevol} we recall basic definitions and we derive a simplified form of the evolution equations best suited for our analysis. In Sec.~\ref{spinorial} we exhibit the underlying $O(3)$ symmetry that enables us to express the solutions of the evolution equations in terms of spinors. The method of spinorial decomposition is extended in this section to cover a time-dependent field. In Sec.~\ref{exp}, using this method, we find an analytic solution of the spinorial equations when the electric field is turned on exponentially. In Sec.~\ref{evol} we use this solution to determine the time evolution of the vacuum. In Sec.~\ref{pert} we show the the consecutive approximations to the analytic solution can also be obtained by perturbation theory. In Sec.~\ref{saut} we construct the analytic solution from the solutions of the Dirac equation found by Sauter \cite{sauter}. Finally, in Sec.~\ref{fin} we compare the time evolution of the QED vacuum in both cases underscoring the main features.

\section{Evolution of the DHW function\\in a time-varying electric field}\label{secevol}

The DHW function (which in fact is a $4\times4$ hermitian matrix) is defined as follows \cite{bgr}:
\begin{eqnarray}\label{dhw}
\fl W_{\alpha\beta}&(\bm{r},\bm{p},t)=-{\textstyle\frac{1}{2}}\int \!\rmd^3\eta\,\rme^{-\rmi \left(\bm{p}\cdot \bm{\eta}/\hbar + \varphi\right)}\left\langle\Phi\left|\left[\hat{\Psi}_{\alpha}
\left(\bm{r}+\bm{\eta}/2,t\right),
\hat{\Psi}_{\beta}^{\dagger}
\left(\bm{r}-\bm{\eta}/2,t\right)\right]\right|\Phi\right\rangle,
\end{eqnarray}
where the square bracket denotes the commutator of the quantized Dirac field operator $\mathbf{\hat{\Psi}}$ with its hermitian conjugate and the phase $\varphi$ is the line integral of the vector potential
\begin{eqnarray}
\varphi=\frac{e}{\hbar}\int_{-1/2}^{1/2}\rmd\lambda\,\bm{\eta}\!\cdot\!\bm{A}
\left(\bm{r}+\lambda\bm{\eta},t\right).
\end{eqnarray}
The physical meaning of this phase factor is best seen by rewriting Eq.~(\ref{dhw}) with the use of the gauge covariant shift operator
\begin{eqnarray}\label{dhw1}
\fl W_{\alpha\beta}&(\bm{r},\bm{p},t)=-{\textstyle\frac{1}{2}}\int \!\rmd^3\eta\,\rme^{-\rmi\bm{p}\cdot \bm{\eta}/\hbar}\left\langle\Phi\left|\left[\rme^{\bm\eta/2\cdot\bm D}\hat{\Psi}_{\alpha}\left(\bm{r},t\right),\rme^{-\bm\eta/2\cdot\bm D}\hat{\Psi}_{\beta}^{\dagger}\left(\bm{r},t\right)\right]\right|\Phi\right\rangle,
\end{eqnarray}
where $\bm D={\bm\nabla}-ie/\hbar{\bm A}(\bm r,t)$. The equivalence of this form of $\bm W$ and the previous one follows from the relations
\begin{eqnarray}\label{rel}
\rme^{\pm\bm\eta/2\cdot\bm D}&=\exp\left[-\rmi\frac{e}{\hbar}\int_{0}^{1/2}\!\rmd\lambda\,\bm{\eta}\!\cdot\!\bm{A}
\left(\bm{r}\pm\lambda\bm{\eta},t\right)\right]\rme^{\pm\bm\eta/2\cdot\bm\nabla}.
\end{eqnarray}

The phase factor has been introduced in \cite{bgr} to secure gauge invariance of the DHW function. That is why only the electromagnetic fields and not the potentials appear in the evolution equations. This also means that the argument $\bm p$ of the DHW function is the kinetic not the canonical momentum.

The most general DHW function has 16 real components. The $4\times 4$ matrix $W_{\alpha\beta}$ can be decomposed into 16 matrices \cite{bgr}. We found it convenient to use the matrices introduced originally by Dirac \cite{dir},
\begin{eqnarray}\label{decomp}
{\bm W}(\bm r,\bm p,t)={\textstyle\frac{1}{4}}\bigl[f_0+\sum_{i=1}^3 \rho_i f_i + \bm\sigma\!\cdot\!{\bm g}_0 + \sum_{i=1}^3 \rho_i \bm\sigma\!\cdot\!{\bm g}_i\bigr].
\end{eqnarray}
In the special case when the initial state is the vacuum state and the applied electric field is spatially uniform, only the following quantities do not vanish: the mass density $f_3(\bm r,\bm p,t)$, the spin density $\bm g_0(\bm r,\bm p,t)$, the electric current density $\bm g_1(\bm r,\bm p,t)$, and the magnetic moment density $\bm g_2(\bm r,\bm p,t)$. Moreover, as shown in \cite{bbr,bgr}, only three independent coefficients $(w_1,w_2,w_3)$ are needed to express all these quantities, namely
\begin{eqnarray}\label{10to3}
-\frac{1}{2}\left[\begin{array}{c}f_3\\\bm g_0\\\bm g_1\\\bm g_2\end{array}\right]
\!=\frac{w_1}{E_\perp}\!\!\left[\begin{array}{c}mc^2\\0\\ c\bm p_\perp\\0\end{array}\right]
\!+\frac{w_2}{E_\perp}\!\!\left[\begin{array}{c}0\\c\bm p_\perp\times\bm n\\0\\mc^2\bm n\end{array}\right]
\!+w_3\!\!\left[\begin{array}{c}0\\0\\\bm n\\0\end{array}\right],
\end{eqnarray}
where $\bm n$ is the unit vector in the direction of the electric field, $\bm p_\perp$ is the component of $\bm p$ perpendicular to the field, and $E_\perp$ is the transverse energy, $E_\perp=\sqrt{m^2c^4+c^2p_\perp^2}$. The factor -1/2 has been introduced to have the vector $(w_1,w_2,w_3)$ normalized to 1.

The evolution equations for the three coefficients were derived in \cite{bbr} and they are:
\begin{eqnarray}\label{eqmw}
\left(\partial_\tau+\mathcal{E}(\tau)\partial_q\right)
\left[\begin{array}{c}w_1\\w_2\\w_3\end{array}\right]
=2\left[\begin{array}{ccc}0&q&0\\-q&0&1\\0&-1&0\end{array}\right]
\left[\begin{array}{c}w_1\\w_2\\w_3\end{array}\right].
\end{eqnarray}
These equations differ from Eqs.~(11) of Ref.~\cite{bbr} only by the time-dependence of the electric field. We introduced here the dimensionless variables, namely
\begin{eqnarray}
\tau=\frac{E_\perp t}{\hbar},\quad q=\frac{cp_\parallel}{E_\perp},\quad
\mathcal{E}(\tau)=\frac{e\hbar c E(\tau)}{E_\perp^2},
\end{eqnarray}
where $p_\parallel$ is the component of ${\bm p}$ in the direction of the electric field.

We shall expose the group-theoretic content of the evolution equations for the vector ${\bm w}(q,\tau)$ by writing them with the use of the spin-one matrices ${\bm S}$ in the Cartesian basis,
\begin{eqnarray}\label{eqmw1}
\left(\partial_\tau+\mathcal{E}(\tau)\partial_q\right){\bm w}(q,\tau)
=2\rmi{\bm S}\!\cdot\!{\bm{\mathcal{H}}}(q){\bm w}(q,\tau),
\end{eqnarray}
where
\begin{eqnarray}\label{s}
{\bm{\mathcal{H}}}(q)=(1,0,q),\quad{\bm S}\!\cdot\!{\bm{\mathcal{H}}}(q)
=-\rmi\left[\begin{array}{ccc}0&q&0\\-q&0&1\\0&-1&0
\end{array}\right].
\end{eqnarray}
In this way we may interpret the evolution as a rotation in the three-dimensional vector space \cite{err1}.

In order to convert partial differential equations into ordinary differential equations we use an extension of the method of characteristics to systems of equations. However, this time the conversion goes in opposite direction as compared to our previous work. Namely, we shall eliminate the derivative with respect to $q$, not the time derivative. This is achieved by replacing the vector ${\bm w}(q,\tau)$ by a new vector ${\bm{\mathcal X}}(q,\tau)$ with the momentum variable $q$ displaced by the vector potential ${\mathcal A}(\tau)=e\hbar c A(\tau)/E_\perp^2$,
\begin{eqnarray}
q\mapsto q+{\mathcal A}(\tau),\quad {\mathcal A}(\tau)=-\int_{\tau_0}^{\tau}\rmd\tau'\mathcal{E}(\tau').
\end{eqnarray}
Formally, this is accomplished by the following substitution:
\begin{eqnarray}\label{subst}
{\bm w}(q,\tau)=\exp\left[{\mathcal A}(\tau)\partial_q\right]{\bm{\mathcal X}}(q,\tau).
\end{eqnarray}
When this expression is inserted into Eqs.~(\ref{eqmw1}) we find that the components of this new vector satisfy ordinary differential equations with $q$ playing the role of a fixed parameter,
\begin{eqnarray}\label{eqmx}
\frac{\rmd{\bm{\mathcal X}}(q,\tau)}{\rmd\tau}
=2\rmi{\bm S}\!\cdot\!{\bm{\mathcal{H}}}(q-{\mathcal A}(\tau)){\bm{\mathcal X}}(q,\tau).
\end{eqnarray}
This equation was obtained with the use of the relation
\begin{eqnarray}\label{rel1}
\exp\left[-{\mathcal A}(\tau)\partial_q\right]{\bm{\mathcal{H}}}(q)
\exp\left[{\mathcal A}(\tau)\partial_q\right]={\bm{\mathcal{H}}}(q-{\mathcal A}(\tau)).
\end{eqnarray}
The solution of the initial value problem will proceed by construction of the fundamental matrix $\hat{\mathcal F}(q,\tau)$ for the set of equations (\ref{eqmx}). This matrix is built (cf., for example, \cite{gt}) from a set of three linearly independent solutions $({\bm{\mathcal X}}_1(q,\tau),{\bm{\mathcal X}}_2(q,\tau),{\bm{\mathcal X}}_3(q,\tau))$ of Eqs.~(\ref{eqmx}),
\begin{eqnarray}\label{matw}
\hat{\mathcal F}(q,\tau)=\left[\begin{array}{ccc}{\mathcal{X}}_1^1&{\mathcal{X}}_2^1&{\mathcal{X}}_3^1\\
{\mathcal{X}}_1^2&{\mathcal{X}}_2^2&{\mathcal{X}}_3^2\\
{\mathcal{X}}_1^3&{\mathcal{X}}_2^3&{\mathcal{X}}_3^3
\end{array}\right].
\end{eqnarray}
The fundamental matrix normalized to the unit matrix at $\tau_0$ is
\begin{eqnarray}\label{matnorm}
\hat{\mathcal F}_N(q,\tau,\tau_0)=\hat{\mathcal F}(q,\tau)\hat{\mathcal F}^{-1}(q,\tau_0).
\end{eqnarray}
The solution of the initial value problem for the vector ${\bm{\mathcal X}}$ with an arbitrary initial condition ${\bm{\mathcal X}}(q,\tau_0)$ set at $\tau=\tau_0$ is therefore
\begin{eqnarray}\label{finx}
{\bm{\mathcal X}}(q,\tau)=\hat{\mathcal F}_N(q,\tau,\tau_0){\bm{\mathcal X}}(q,\tau_0).
\end{eqnarray}

Our aim is to find the time evolution of the vacuum state. The DHW function for this state is built from the following set of $w_i$ functions \cite{bgr}:
\begin{eqnarray}\label{vac}
w_1({\bm p})=\frac{E_\perp}{E_p},\quad w_2({\bm p})=0,\quad
w_3({\bm p})=\frac{c{\bm p}\!\cdot\!{\bm n}}{E_p},
\end{eqnarray}
where $E_p=\sqrt{m^2c^4+c^2p^2}$ is the total energy.
Thus, in our dimensionless units the vector describing the vacuum has the form:
\begin{eqnarray}\label{vacuum}
{\bm V}(q)=\frac{1}{E_q}(1,0,q),
\end{eqnarray}
where $E_q=\sqrt{1+q^2}$ represents the dimensionless total energy. Choosing ${\bm{\mathcal X}}(q,\tau_0)={\bm V}(q)$ as the initial state we obtain the time evolved DHW function by inserting the formula (\ref{finx}) into Eq.~(\ref{subst})
\begin{eqnarray}\label{finw}
{\bm{w}}(q,\tau)=\hat{\mathcal F}_N(q+{\mathcal A}(\tau),\tau,\tau_0){\bm{V}}(q+{\mathcal A}(\tau)).
\end{eqnarray}
All we need now to solve the initial value problem is the fundamental matrix. In the next section we show how to construct this matrix using the method of spinorial decomposition.

\section{Spinorial decomposition}\label{spinorial}

The set of three differential equations leads to third order equations for each component studied in Ref.~\cite{torado}. As we have shown in \cite{bbr}, one may replace third order equations with second order equations by employing the method of spinorial decomposition that exploits the $O(3)$ group symmetry. We shall extend this method here to cover the case of the time-varying field. To this end, we introduce the following representation for the components of the vector $\bm{\mathcal X}$ in terms of a two-component spinor $\Psi$ and the Pauli matrices,
\begin{eqnarray}\label{wpsi}
{\mathcal X}^1=\Psi^\dagger\sigma_x\Psi,
\quad{\mathcal X}^2=\Psi^\dagger\sigma_y\Psi,
\quad{\mathcal X}^3=\Psi^\dagger\sigma_z\Psi.
\end{eqnarray}
One may check by inspection that the evolution equations (\ref{eqmx}) for the vector $\bm{\mathcal X}$ follow from the evolution equations for the spinor $\Psi$ and its conjugate,
\numparts
\begin{eqnarray}
\frac{\rmd\Psi(q,\tau)}{\rmd\tau}&=\rmi{\bm\sigma}\!\cdot\!
{\bm{\mathcal{H}}}(q-{\mathcal A}(\tau))\Psi(q,\tau),\label{eqmp1}\\
\frac{\rmd\Psi^\dagger(q,\tau)}{\rmd\tau}&=-\rmi\Psi^\dagger(q,\tau){\bm\sigma}\!\cdot\!
{\bm{\mathcal{H}}}(q-{\mathcal A}(\tau)).
\end{eqnarray}
\endnumparts
These two coupled equations
\numparts
\begin{eqnarray}
\frac{\rmd\psi_-(q,\tau)}{\rmd\tau}&=\rmi(q-{\mathcal A}(\tau))\psi_-(q,\tau)+\rmi\psi_+(q,\tau),\label{firstord}\\
\frac{\rmd\psi_+(q,\tau)}{\rmd\tau}&=-\rmi(q-{\mathcal A}(\tau))\psi_+(q,\tau)+\rmi\psi_-(q,\tau)
\end{eqnarray}
\endnumparts
can be transformed to the following two separate second order equations for the upper $\psi_-$ and lower $\psi_+$ components of the spinor,
\begin{eqnarray}\label{secord}
\left[\frac{\rmd^2}{\rmd\tau^2}+1+(q-{\mathcal A}(\tau))^2
\mp \rmi{\mathcal{E}}(\tau)\right]\psi_\mp(q,\tau)=0.
\end{eqnarray}
These equations coincide with the second order equations obtained from the Dirac equation in the time-dependent potential ${\mathcal A}(\tau)$ along the $z$ axis after the separation of the spatial variables through the following substitution
\begin{eqnarray}\label{dir}
\psi({\bm r},t)=\exp(\rmi{\bm p}\cdot{\bm r})\psi(t).
\end{eqnarray}
Therefore, known analytic solutions of the reduced Dirac equation (\ref{secord}) will lead to analytic solutions for the Wigner function. Linearly independent solutions of Eqs.~(\ref{secord}) serve as building blocks in the construction of the spinors from which we shall construct the fundamental matrix (\ref{matnorm}). The most convenient choice of two spinorial solutions is when they are orthonormal. Let us suppose that we found one solution $\Psi_1(q,\tau)$ of Eqs.~(\ref{eqmp1}-{\it b}),
\begin{eqnarray}\label{first}
\Psi_1(q,\tau)=\left(\begin{array}{c}f\\-g\end{array}\right).
\end{eqnarray}
The minus sign in the lower component is chosen for future convenience. Then, the second solution of Eqs.~(\ref{eqmp1}-{\it b}) can always be chosen in the form
\begin{eqnarray}\label{second}
\Psi_2(q,\tau)=\left(\begin{array}{c}g^*\\f^*\end{array}\right).
\end{eqnarray}
It is automatically orthogonal to the first solution and it is normalized if the first solution was normalized.

The following formulas, introduced in \cite{bbr}, show how to produce three orthonormal vectors ${\bm{\mathcal{X}}}_k$ from two orthonormal spinors $\Psi_1$ and $\Psi_2$:
\numparts
\begin{eqnarray}
{\bm{\mathcal{X}}}_1&=-\frac{1}{2}\left(\Psi_1^\dagger\bm{\sigma}\Psi_2
+\Psi_2^\dagger\bm{\sigma}\Psi_1\right),\label{3vec}\\
{\bm{\mathcal{X}}}_2&=-\frac{1}{2}\left(-\rmi\Psi_1^\dagger\bm{\sigma}\Psi_2
+\rmi\Psi_2^\dagger\bm{\sigma}\Phi_1\right),\\
{\bm{\mathcal{X}}}_3&=-\frac{1}{2}\left(\Psi_1^\dagger\bm{\sigma}\Psi_1
-\Psi_2^\dagger\bm{\sigma}\Psi_2\right).
\end{eqnarray}
\endnumparts
The orthonormality of these vectors is a result of the following identity valid for every set of four spinors,
\begin{eqnarray}\label{id}
\left(\Psi_1^\dagger{\bm\sigma}\Psi_2\right)\!\cdot\!
\left(\Psi_3^\dagger{\bm\sigma}\Psi_4\right)=2\left(\Psi_1^\dagger\Psi_4\right)\left(\Psi_3^\dagger\Psi_2\right)
-\left(\Psi_1^\dagger\Psi_2\right)\left(\Psi_3^\dagger\Psi_4\right).
\end{eqnarray}
The fundamental matrix built from orthonormal vectors (\ref{3vec}-{\it c}) is orthogonal. Therefore, in the formula (\ref{matnorm}) we may replace the inverse matrix by the transposed matrix. We have chosen the minus sign in the definition (\ref{3vec}-{\it c}) to have the vectors $({\bm{\mathcal{X}}}_1,{\bm{\mathcal{X}}}_2,{\bm{\mathcal{X}}}_3)$ forming a right-handed set, ${\bm{\mathcal{X}}}_1\!\cdot({\bm{\mathcal{X}}}_2\times{\bm{\mathcal{X}}}_3)=1$, so that the determinant of the fundamental matrix is equal to 1 ($\hat{\mathcal F}$ describes a pure rotation).

We may now use the three orthonormal vectors (\ref{3vec}-{\it c}) to build the matrix ${\hat{\mathcal F}}$. The components of this matrix \cite{err2} are real and imaginary parts of various quadratic expressions of the two functions $f$ and $g$,
\begin{eqnarray}\label{3in2}
{\hat{\mathcal F}}=\left[\begin{array}{ccc}-\Re[f^2-g^2]&\Im[f^2-g^2]&2\Re[fg^*]\\
\Im[f^2+g^2]&\Re[f^2+g^2]&-2\Im[fg^*]\\
-2\Re[fg]&2\Im[fg]&gg^*-ff^*\end{array}\right].
\end{eqnarray}
We need a specific form of the function ${\mathcal A}(\tau)$ to proceed any further. This will be done in next sections.

\section{Electric field switched on at a constant rate}\label{exp}

The adiabatic switching on of the electric field will be described by an exponential prefactor. Thus, the electric field will be given by the formula ${\bm{\mathcal{E}}}(\tau)={\mathcal{E}}\exp(b\tau){\bm n}$ with $b>0$. It follows from the Maxwell equations that this field is produced by the current $b{\bm{\mathcal{E}}}(\tau)$. For sufficiently small values of $b$ this current can be neglected in comparison with the current associated with the produced pairs.

Assuming that the initial time $\tau_0$ lies in the remote past ($\tau_0=-\infty$), we obtain
\begin{eqnarray}\label{q}
{\mathcal A}(\tau)=-{\mathcal{E}}b^{-1}\exp(b\tau).
\end{eqnarray}
This choice of the time dependence results in the following second order equations (\ref{secord}) for the upper and lower spinor components:
\begin{eqnarray}\label{secord1}
\left[ \frac{\rmd^2}{\rmd\tau^2}+1+(q+\frac{\mathcal{E}}{b}\rme^{b\tau})^2
\mp \rmi{\mathcal{E}}\rme^{b\tau} \right] \psi_\mp(q,\tau)=0.
\end{eqnarray}
These equations are solvable in terms of confluent hypergeometric functions. The two independent solutions for $\psi_-$ have the form
\numparts
\begin{eqnarray}
&f(q,\tau)=\frac{\exp(\rmi\tau E_q)\rme^{\rmi s/2}}
{\sqrt 2\sqrt{E_q^2-qE_q}}~
{_{1\!}F_1}\left(\rmi\frac{E_q-q}{b},1+2\rmi\frac{E_q}{b};-\rmi s\right),\label{2sol}\\
&g^*(q,\tau)=-\frac{\exp(-\rmi\tau E_q)\rme^{-\rmi s/2}}
{\sqrt 2\sqrt{E_q^2+qE_q}}~
{_{1\!}F_1}\left(1-\rmi\frac{E_q-q}{b},1-2\rmi\frac{E_q}{b};\rmi s\right),
\end{eqnarray}
\endnumparts
where $s=2{\mathcal{E}}b^{-2}\rme^{b\tau}$. The lower components are, in general, linear combinations of both $f^*(q,\tau)$ and $g(q,\tau)$. However, for our choice (\ref{2sol}-{\it b}) of the upper components the lower components $\psi_+$ contain only one term and the two orthonormal solutions of Eqs.~(\ref{eqmp1}-{\it b}) are (\ref{first}) and (\ref{second}). This is the simplest pair of solutions because each component contains only one confluent hypergeometric function. Orthogonality of the two spinors is self-evident but the check of normalization requires the use of the evolution equations (\ref{eqmp1}-{\it b}). Since these equations preserve scalar products of spinors, it is sufficient to check the normalization at any value of $\tau$ and we choose $\tau=-\infty$ or $s=0$. At this value, the confluent hypergeometric functions reduce to 1 and the normalization $|f|^2+|g|^2=1$ follows immediately. It may be worth noticing that the normalization condition for an arbitrary time generates some quadratic identities for the confluent functions that seem to be unknown.

To elucidate the connection between the results obtained for the time-independent field in \cite{bbr} and the time-dependent field in the present case, we concoct one ``super'' equation that encompasses these two cases. The evolution equation (\ref{eqmp1}-{\it b}) will have the form:
\begin{eqnarray}
\frac{\rmd\Psi(q,\tau)}{\rmd\tau}&=\rmi{\bm\sigma}\!\cdot\!
{\bm{\mathcal{H}}}\left(q+{\mathcal E}\frac{\exp(b\tau)-\exp(b\tau_0)}{b}\right)\Psi(q,\tau),
\end{eqnarray}
where $\tau_0$ specifies the initial moment at which $\Psi$ describes the field-free vacuum. The exact solution of this equation can still be found. All we have to do is to replace $q$ by $q-\exp(b\tau_0)/b$ in (\ref{2sol}-{\it b}). There are two {\em mutually exclusive} limiting procedures that lead either to the solution of Eqs.~(22) in \cite{bbr} or to the time-dependent solution given by (\ref{2sol}-{\it b}). The first limit is when $\tau_0=0$ and $b=0$ and it leads to a solution singular at $\mathcal E=0$. This solution reproduces the Schwinger singularity as shown in \cite{bbr}. The second limit is when $\tau_0 \to -\infty$. This solution is analytic in $\mathcal E$ but it has a singularity when $b \to 0$. Thus, there is no direct connection between these two limiting cases.

\section{Time evolution of the QED vacuum}\label{evol}

One must be careful with taking the limit $\tau_0\to -\infty$ in Eq.~(\ref{finw}) because of the oscillatory terms present in the fundamental matrix even in the limit of vanishing field. However, these terms cancel out when the initial state is the vacuum,
\begin{eqnarray}\label{cancel}
\lim_{\tau_0\to -\infty}\hat{\mathcal F}^{T}(q,\tau_0){\bm{V}}(q)=(0,0,-1).
\end{eqnarray}
The final expression for the vector ${\bm w}_{\rm v}(q,\tau)$ that evolved from the vacuum at $\tau_0=-\infty$ is obtained by combining the formulas (\ref{finw}), (\ref{3in2}), and (\ref{cancel}),
\begin{eqnarray}\label{finv}
\fl {\bm w}_{\rm v}(q,\tau)&=\hat{\mathcal F}(q+{\mathcal A}(\tau),\tau)(0,0,-1)=\left(-2\Re({\tilde f}{\tilde g}^*),2\Im({\tilde f}{\tilde g}^*),|{\tilde f}|^2-|{\tilde g}|^2\right),
\end{eqnarray}
where ${\tilde f}=f(q+{\mathcal A}(\tau),\tau)$ and ${\tilde g}=g(q+{\mathcal A}(\tau),\tau)$. This vector satisfies the evolution equations (\ref{eqmw}) and the vacuum initial condition. The validity of these statements follows from our general procedure but it can also be checked by a direct (albeit tedious) calculation.

The density of pairs $n(q,\tau)$ produced by the electric field \cite{err3} was shown in \cite{bgr} to be
\begin{eqnarray}\label{n}
n(q,\tau)=1-O(q,\tau),
\end{eqnarray}
where $O(q,\tau)$ is the overlap between the DHW vector at time $\tau$ given by (\ref{finv}) and the initial vacuum vector (\ref{vacuum}),
\begin{eqnarray}\label{ovlp}
O(q,\tau)&={\bm V}(q)\!\cdot\!{\bm w}_{\rm v}(q,\tau)=\frac{q(|{\tilde f}|^2-|{\tilde g}|^2)-2\Re[{\tilde f}{\tilde g}^*]}{E_q}.
\end{eqnarray}
Since both vectors ${\bm V}(q)$ and ${\bm w}_{\rm v}(q,\tau)$ are normalized, $1-O(q,\tau)$ is simply one-half of the squared difference between the initial and the evolved vectors.
The relation (\ref{n}) has been guessed from numerical calculations in Ref.~\cite{bgr}. In the Appendix we give a derivation of this important formula and show that it is valid for an arbitrary electromagnetic field. In the present case $f_0$ vanishes because the charge density in a state that evolved from the vacuum under the action of a uniform field vanishes at all times. Therefore, the general formula (\ref{nofel2}) reduces to
\begin{eqnarray}\label{nofel3}
n({\bm r},{\bm p},t)=\left[1+\frac{m f_3({\bm r},{\bm p},t)+{\bm p}\!\cdot\!{\bm g}_1({\bm r},{\bm p},t)}{2E_p}\right].
\end{eqnarray}
With the use of Eqs.~(\ref{10to3}) and (\ref{vac}) the second term can be identified with (minus) the overlap function and the formula (\ref{n}) is thus proven.
\begin{figure}
\centering
\includegraphics[scale=0.45]{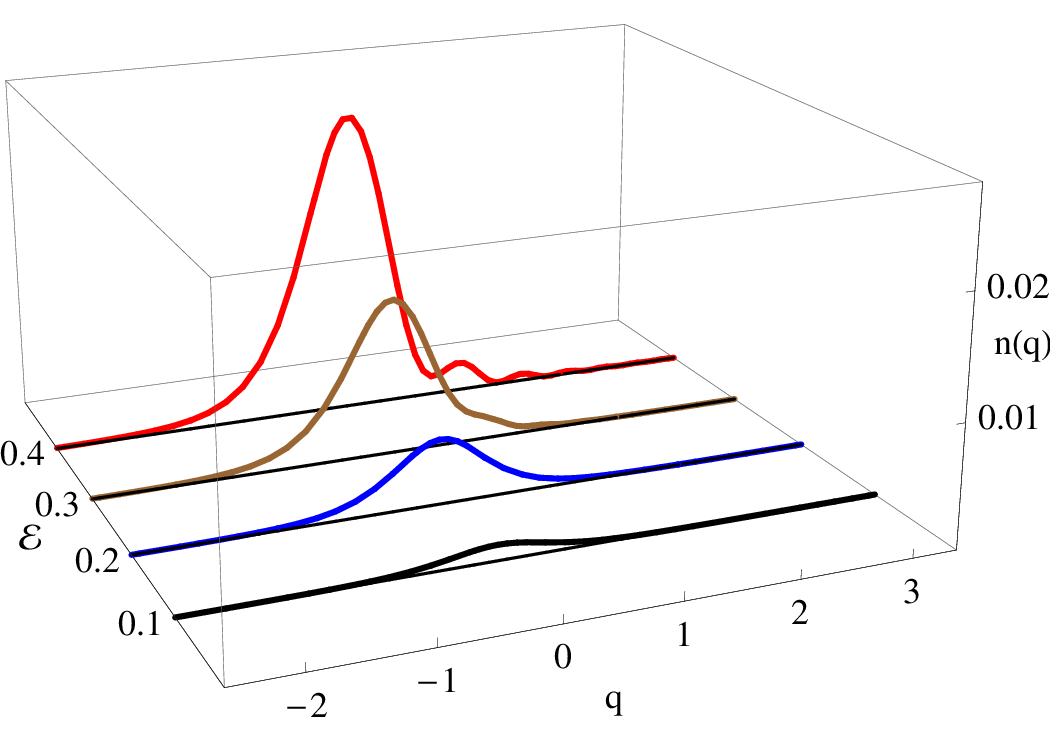}
\caption{The pair density $n(q,\tau)$ for the exponentially growing field at time $\tau=0$ and for $b=0.05$ plotted as a function of the dimensionless momentum $q$ at the values 0.1, 0.2, 0.3, and 0.4 of the dimensionless field strength $\mathcal{E}$.}\label{fig1}
\end{figure}

The number of pairs should not change under the simultaneous reversal of the field and the parallel momentum because it is only the relative orientation of these two vectors that matters. This property directly follows from the fact that under the change of signs ${\mathcal{E}}\to-{\mathcal{E}},\;q\to-q$ the functions ${\tilde f}$ and ${\tilde g}$ are interchanged ${\tilde f}\to-{\tilde g},\;{\tilde g}\to-{\tilde f}$.

In Fig.~\ref{fig1} we show the pair density as a function of $q$ for different values of the field strength $\mathcal{E}$. In contrast to the results of Ref.~\cite{bbr}, there are no transients. The number of pairs is a smooth function of $q$.

The most striking result is the absence of an essential singularity in the dependence of the number of produced pairs on the field strength $\mathcal{E}$. All functions appearing in the formula (\ref{ovlp}), as long as $b\neq0$, can be expanded into a convergent power series in $\mathcal{E}$ because the confluent hypergeometric function $_1F_1(a,c;z)$ is an analytic function of its three arguments. However, it has poles in $c$ at all negative integers $-n$ (including 0). Since in the formulas (\ref{2sol}-{\it b}) the parameter $c$ is a function of $\mathcal{E}$,
\begin{eqnarray}\label{c}
c=1\pm 2\rmi\frac{\sqrt{1+(q-\mathcal{E}/b)^2}}{b},
\end{eqnarray}
the radius of convergence $R$ of the series in $\mathcal{E}$ can be determined from the conditions $c=0$ and it reads
\begin{eqnarray}\label{rad}
R=b\sqrt{1+q^2+b^2/4}.
\end{eqnarray}
Thus, the perturbative expansion is convergent only for $\mathcal{E}$ smaller than the radius $R$. In the limiting case, when $b\to 0$, the radius of convergence goes to zero as it must because of the essential singularity that reappears in this limit.
\begin{figure}
\centering
\includegraphics[scale=0.45]{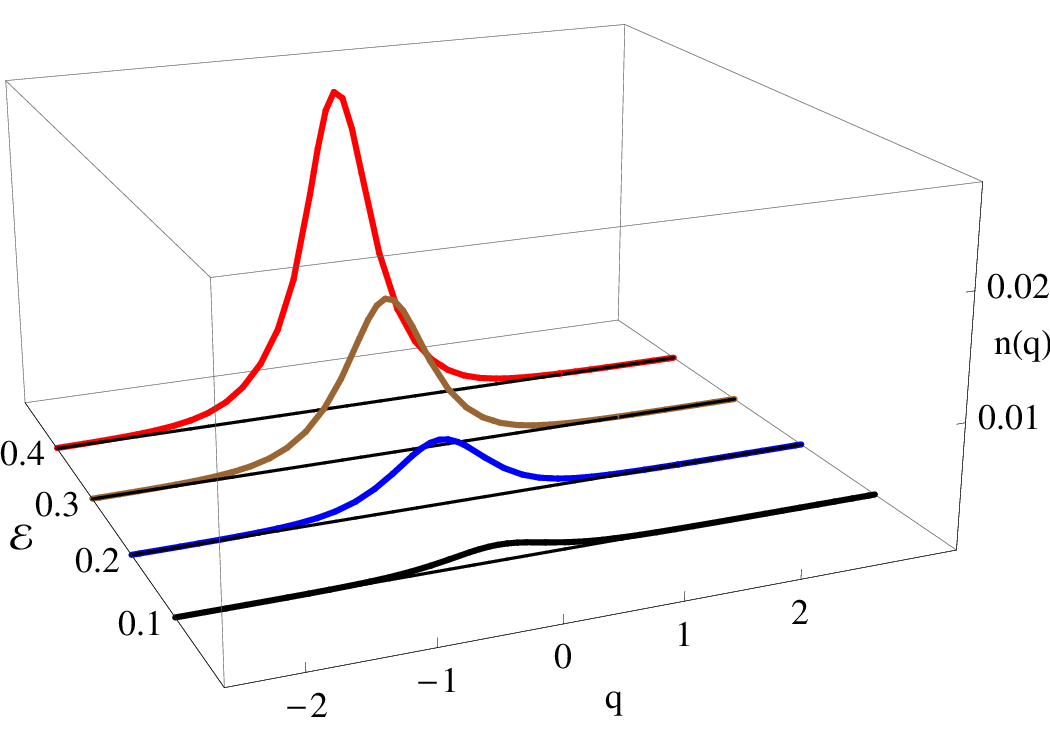}
\caption{The analogous plot as in Fig.~1 obtained from the sixth-order perturbative expansion.}\label{fig2}
\end{figure}

The expansion in $\mathcal{E}$ up to the sixth order for $\tau=0$ gives the following formula:
\begin{eqnarray}\label{six}
\fl O(q,0)=&1-\frac{N_2\mathcal{E}^2}{E_q^6(4+{\tilde b}^2)}-\frac{N_3\mathcal{E}^3}{E_q^{9}(4+{\tilde b}^2)^2(1+{\tilde b}^2)}
-\frac{N_4\mathcal{E}^4}{E_q^{12}(4+{\tilde b}^2)^3(1+{\tilde b}^2)^2(4+9{\tilde b}^2)}\nonumber\\
 \fl &-\frac{N_5\mathcal{E}^5}{E_q^{15}(4+{\tilde b}^2)^4(1+{\tilde b}^2)^3(4+9{\tilde b}^2)^2(1+4{\tilde b}^2)}\nonumber\\
\fl &-\frac{N_6\mathcal{E}^6}{E_q^{18}(4+{\tilde b}^2)^5(1+{\tilde b}^2)^4(4+9{\tilde b}^2)^3(1+4{\tilde b}^2)^2(4+25{\tilde b}^2)}+\mathcal{O}\left(\mathcal{E}^{8}\right),
\end{eqnarray}
where ${\tilde b}=b/E_q$ and
\numparts
\begin{eqnarray}
N_2&=1/2,\label{six1}\\
N_3&=q(7{\tilde b}+{\tilde b}^3),\\
N_4&=3/8\Big(-448q^2+112+{\tilde b}^2(1760q^2-104)+{\tilde b}^4(2164q^2-313)\nonumber\\
&+{\tilde b}^6(472q^2-106)+{\tilde b}^8(36q^2-9)\Big),\\
N_5&=q/2\Big({\tilde b}(-126720 q^2+65472)-{\tilde b}^3(110528 q^2-180048)\nonumber\\
&+{\tilde b}^5(755440 q^2-39900)+{\tilde b}^7(1332348 q^2-425793)\nonumber\\
&+{\tilde b}^9(795136 q^2-379041)+{\tilde b}^{11}(193144 q^2-124383)\nonumber\\
&+{\tilde b}^{13}(24804 q^2-17631)+{\tilde b}^{15}(1296 q^2-972)\Big),\\
N_6&=1/16\Big(48660480 q^4-50282496 q^2+3559424\nonumber\\
&+{\tilde b}^2(-1305722880 q^4+924008448 q^2-30885888)\nonumber\\
&+{\tilde b}^4(-11614668800 q^4+11574500352 q^2-593933568)\nonumber\\
&+{\tilde b}^6(-24204609536 q^4+41551564800 q^2-2746675968)\nonumber\\
&+{\tilde b}^8(21767894400 q^4+52683929664 q^2-5392492368)\nonumber\\
&+{\tilde b}^{10}(151238728256 q^4-21806265408 q^2-3779458248)\nonumber\\
&+{\tilde b}^{12}(236416765448 q^4-131339980524 q^2+2556452997)\nonumber\\
&+{\tilde b}^{14}(183464805216 q^4-150072654336 q^2+6447749532)\nonumber\\
&+{\tilde b}^{16}(78987949264 q^4-83787726072 q^2+4744001262)\nonumber\\
&+{\tilde b}^{18}(19825772704 q^4-25548039744 q^2+1733216492)\nonumber\\
&+{\tilde b}^{20}(3204676008 q^4-4441116924 q^2+339299253)\nonumber\\
&+{\tilde b}^{22}(293855040 q^4-426785760 q^2+34399080)\nonumber\\
&+{\tilde b}^{24}(11664000 q^4-17496000 q^2+1458000)\Big).
\end{eqnarray}
\endnumparts
In Fig.~2 we show that the six terms present in (\ref{six}) reproduce reasonably well the exact results. The structure of the denominators in this formula is a direct consequence of the series expansion of the confluent hypergeometric function
\begin{eqnarray}
_1F_1(a,c;z)=1+\frac{a}{c}\frac{z}{1!}+\frac{a(a+1)}{c(c+1)}\frac{z^2}{2!}+\frac{a(a+1)(a+2)}{c(c+1)(c+2)}\frac{z^3}{3!}+\dots
\end{eqnarray}
In our case $c$ is given by Eq.~(\ref{c}) and the expansion of the expressions $1/(n+c)$ into powers of $\mathcal{E}$ gives
\begin{eqnarray}\label{expe}
\frac{1}{n+c}=\frac{\tilde{b}((n+1)\tilde{b}\mp2\rmi)}{4+(n+1)^2\tilde{b}^2}
\pm\frac{2\rmi q((n+1)\tilde{b}\mp2\rmi)^2}{E_q^3(4+(n+1)^2\tilde{b}^2)^2}\mathcal{E}+\dots.
\end{eqnarray}
Substituting here the consecutive values of $n$ we reproduce the factors in the denominators of (\ref{six}).

Even though each term of the expansion into powers of $\mathcal E$ is finite, for $b=0$ the series becomes only an asymptotic one. In order to make this point absolutely clear, we calculated for $q=0$ the first 8 terms of the perturbation series for the number of produced pairs for two values of $b$.
\begin{eqnarray}\label{8}
\fl n=&\frac{1}{10}\mathcal{E}^{2}-\frac{63}{2600}\mathcal{E}^{4}+\frac{25513}{9802000}\mathcal{E}^{6}
 +\frac{43884005549}{7834150480000}\mathcal{E}^{8}\quad b=1\nonumber\\
\fl &
-\frac{235474684928440509}{26610807325949600000}\mathcal{E}^{10}+\frac{1776480040832261689003353}{180781584001131645584000000}\mathcal{E}^{12}&\nonumber\\
\fl &
-\frac{189228367844501729409089588397558027}{19653421122106203692350489578400000000}\mathcal{E}^{14}&\nonumber\\
\fl &+\frac{171936287478376823113691752600797604830985578228669}{19571206822106176531328917776928027793134400000000000}
\mathcal{E}^{16}+\mathcal{O}\left(\mathcal{E}^{18}\right)\!,&\\
\fl n=&\frac{1}{8}\mathcal{E}^{2}+\frac{21}{128}\mathcal{E}^{4}+\frac{869}{1024}\mathcal{E}^{6}
+\frac{334477}{32768}\mathcal{E}^{8}+\frac{59697183}{262144}\mathcal{E}^{10}\quad b=0\nonumber\\
\fl &+\frac{34429291905}{4194304}\mathcal{E}^{12} +\frac{14631594576045}{33554432}\mathcal{E}^{14} +\frac{68787420596367165}{2147483648}\mathcal{E}^{16}+\mathcal{O}\left(\mathcal{E}^{18}\right)\!.&
\end{eqnarray}
These expansions leave no doubt that the first series is convergent, at least for $\mathcal{E}<1$ since the coefficients tend to zero. The convergence is further improved due to the alternating signs of the coefficients. The second series is only asymptotic since the coefficients grow exponentially and they are all positive. This was to be expected because the case $b=0$ corresponds to the time-independent electric field when the Schwinger formula precludes the possibility of an expansion in powers of $\mathcal{E}$. We shall illustrate these results with the graphs of the pair density.

\section{Perturbation theory}\label{pert}

The solution of the evolution equation (\ref{eqmx}) for the vector ${\bm{\mathcal X}}(q,\tau)$ in perturbation theory is most easily obtained by expanding this vector in the eigenbasis of the matrix appearing in (\ref{eqmw}). This matrix has eigenvalues 0 and $\pm2\rmi E_q$. Its orthonormal eigenvectors are:
\numparts
\begin{eqnarray}\label{ee}
{\bm e}_0(q)&=\frac{1}{E_q}\left[\begin{array}{c}1\\0\\q\end{array}\right],\label{xx}\\
{\bm e}_\pm(q)&=\frac{1}{\sqrt{2}E_q}\left[\begin{array}{c}-q\\\mp \rmi E_q\\1\end{array}\right].
\end{eqnarray}
\endnumparts
We shall seek the solutions of equations (\ref{eqmx}) in the form
\begin{eqnarray}\label{form}
{\bm{\mathcal X}}(q,\tau)&=x_0(q,\tau){\bm e}_0(q)+x_+(q,\tau){\bm e}_+(q)+x_-(q,\tau){\bm e}_-(q).
\end{eqnarray}
The evolution equations for the expansion coefficients are
\numparts
\begin{eqnarray}
\frac{\rmd x_0(q,\tau)}{\rmd \tau}&=-\rmi\sqrt{2}{\tilde {\mathcal A}}(\tau)\left(x_-(q,\tau)-x_+(q,\tau)\right),\label{eqxx}\\
\frac{\rmd x_+(q,\tau)}{d\tau}&=2\rmi E_qx_+(q,\tau)-\rmi{\tilde {\mathcal A}}(\tau)\left(2q\,x_+(q,\tau)-\sqrt{2}x_0(q,\tau)\right),\\
\frac{\rmd x_-(q,\tau)}{\rmd \tau}&=-2\rmi E_qx_-(q,\tau)+\rmi{\tilde {\mathcal A}}(\tau)\left(2q\,x_-(q,\tau)-\sqrt{2}x_0(q,\tau)\right),
\end{eqnarray}
\endnumparts
where ${\tilde {\mathcal A}}(\tau)={\mathcal A}(\tau)/E_q$. In order to prepare these equations for the perturbative solution we shall multiply the second equation by $\exp(-2\rmi E_q\tau)$ and the third equation by $\exp(2\rmi E_q\tau)$ and then integrate all three equation with respect to $\tau$ from $-\infty$ to $\tau$. Assuming that initially at $\tau=-\infty$ the system is in the vacuum state, i.e. $x_0(q,-\infty)=1,\;x_\pm(q,-\infty)=0$, we obtain after integration by parts the following set of three coupled integral equations:
\numparts
\begin{eqnarray}
x_0(q,\tau)&=1-\rmi\sqrt{2}\int_{-\infty}^\tau\!\!\!\!\rmd \tau_1 {\tilde {\mathcal A}}(\tau_1)\left(x_-(q,\tau_1)-x_+(q,\tau_1)\right),\label{int}\\
x_+(q,\tau)&=-\rmi\int_{-\infty}^\tau\!\!\!\!d\tau_1e^{2\rmi E_q(\tau-\tau_1)} {\tilde {\mathcal A}}(\tau_1)\left(2q\,x_+(q,\tau_1)-\sqrt{2}\,x_0(q,\tau_1)\right),\\
x_-(q,\tau)&=\rmi\int_{-\infty}^\tau\!\!\!\!d\tau_1 \rme^{-2\rmi E_q(\tau-\tau_1)} {\tilde {\mathcal A}}(\tau_1)\left(2q\,x_-(q,\tau_1)-\sqrt{2}\,x_0(q,\tau_1)\right).
\end{eqnarray}
\endnumparts
The solution of these equations by iteration produces the perturbative series in powers of the field strength which coincides with the power expansion of the analytic solution. We checked that perturbation theory reproduces the formula (\ref{six}). Note that the function ${\mathcal A}(\tau)$ blows up for $b=0$. Hence, the integral equations (\ref{int}-{\it c}) become invalid which is another manifestation of the Schwinger singularity in the absence of adiabatic switching.

\section{Electric field switched on at a decreasing rate}\label{saut}

To complete our study we shall show now that the main conclusion of our investigation---the removal of the essential singularity---holds also when the time dependence of the electric field is different. To this end we use analytic solutions of the Dirac equation in a uniform time-dependent field discovered by Sauter \cite{sauter} and analyzed in more detail by Sommerfeld \cite{somm}. The time dependence of the electric field in the Sauter solution is ${\bm{\mathcal{E}}}(\tau)={\mathcal{E}}{\rm sech}^2(b\tau/2){\bm n}$ so that ${\mathcal A}(\tau)=-2{\mathcal E}(1+\tanh(b\tau/2))$. In this case, as in the previous one, the electric field is turned on gradually, starting from the zero value at $t_0=-\infty$.

The second order equations (\ref{secord}) for the Sauter field have the following form:
\begin{eqnarray}\label{secord2}
\fl \left[ {\frac{\rmd^2}{\rmd\tau^2}+1+\left(q+\frac{2\mathcal{E}(1+\tanh(b\tau/2))}{b}\right)^2
\mp \rmi\frac{\mathcal{E}}{\cosh^2(b\tau/2)}} \right] \psi_\mp(q,\tau)=0.
\end{eqnarray}
Following Sauter and Sommerfeld we replace the time variable by $u=(1+\tanh(b\tau/2))/2$ and write down the solutions in terms of hypergeometric functions. The first order equations in this new variable read:
\numparts
\begin{eqnarray}
bu(1-u)\frac{\rmd \psi_-(q,u)}{\rmd u}&=\rmi(q+\frac{4{\mathcal E}u}{b})\psi_-(q,u)+\rmi\psi_+(q,u),\label{firstord2}\\
bu(1-u)\frac{\rmd \psi_+(q,u)}{\rmd u}&=-\rmi(q+\frac{4{\mathcal E}u}{b})\psi_+(q,u)+\rmi\psi_-(q,u).
\end{eqnarray}
\endnumparts
Two orthonormal solutions (\ref{first}) and (\ref{second}) of the first order equations (\ref{firstord}-{\it b}) are constructed this time from the following two solutions of the second order equation for $\psi_-$:
\numparts
\begin{eqnarray}
\fl &f(q,u)=\frac{u^{\rmi\alpha}(1-u)^{-\rmi\beta}}{\sqrt 2\sqrt{E_q^2-qE_q}}~{_{2}F_1}\left(4\rmi\frac{\mathcal{E}}{b^2}+\rmi\alpha
-\rmi\beta,1-4\rmi\frac{\mathcal{E}}{b^2}+\rmi\alpha-\rmi\beta,1+2\rmi\alpha;u\right),\label{3sol}\\
\fl &g^*(q,u)=-\frac{u^{-\rmi\alpha}(1-u)^{\rmi\beta}}{\sqrt 2\sqrt{E_q^2+qE_q}}~{_{2}F_1}\left(1-4\rmi\frac{\mathcal{E}}{b^2}-\rmi\alpha+\rmi\beta,4\rmi\frac{\mathcal{E}}{b^2}-\rmi\alpha
+\rmi\beta,1-2\rmi\alpha;u\right),
\end{eqnarray}
\endnumparts
where
\begin{eqnarray}
\alpha=\frac{\sqrt{1+q^2}}{b},\;\beta=\frac{\sqrt{1+(q+4\mathcal{E}/b)^2}}{b}.
\end{eqnarray}
\begin{figure}
\centering
\includegraphics[scale=0.45]{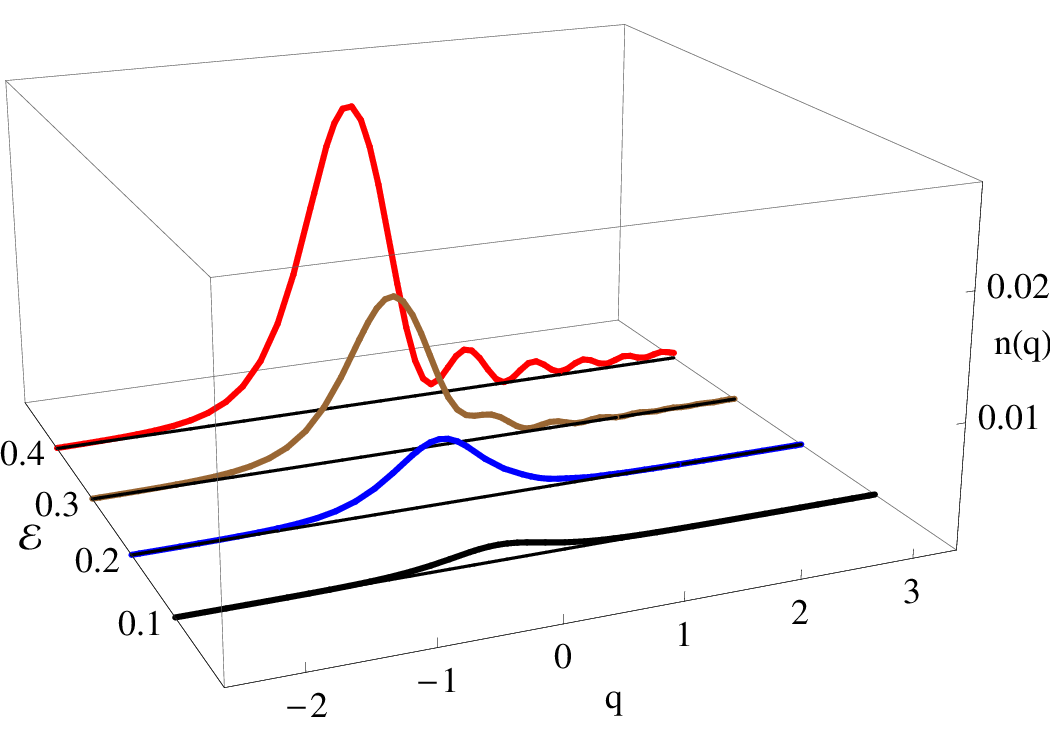}
\caption{The pair density $n(q)$ for the Sauter field at time $\tau=0$ and for $b=0.05$ plotted as a function of the dimensionless momentum $q$ at the values 0.1, 0.2, 0.3, and 0.4 of the dimensionless field strength $\mathcal{E}$.}\label{fig3}
\end{figure}The expression for the number of pairs constructed according to Eqs.~(\ref{n}) and (\ref{ovlp}) is an analytic function of $\mathcal{E}$ but this time the radius of convergence is four times smaller than the value (\ref{rad}) for the exponential field. For the Sauter field we reproduce the formula for the number of produced pairs at $t=+\infty$ that were obtained by Narozhnyi and Nikishov \cite{nn},
\begin{eqnarray}
2\frac{\sinh\left(\frac{4\pi\mathcal{E}}{b^2}
-\pi\left(\alpha-{\bar\beta}\right)\right)\sinh\left(\frac{4\pi\mathcal{E}}{b^2}+\pi  \left(\alpha-{\bar\beta}\right)\right)}{\sinh\left(2\pi\alpha\right)
\sinh\left(2\pi{\bar\beta}\right)},
\end{eqnarray}
where $\bar\beta$ differs from $\beta$ by having the sign of $\mathcal{E}$ reversed.

\section{Common properties of the two solutions}\label{fin}

In \cite{bbr} we derived a general expression for the density of pairs $n(q)$ created by a constant electric field during the finite time interval $\left[\tau_0,\tau\right]$. From this result, in the limit $\tau_0 \rightarrow-\infty$ corresponding to the infinite duration of the pair creation process, we obtain the following time independent expression found in \cite{hag}:
\begin{eqnarray}\label{nqstat}
n(q) = \frac{1}{\sqrt{1+q^2}}\left|f_c(-q)\sqrt{\sqrt{1+q^2}-q}-g_c(-q)\sqrt{\sqrt{1+q^2}+q}\right|^2,
\end{eqnarray}
where
\numparts
\begin{eqnarray}
f_c(q)=\frac{\rme^{-\pi/(8\mathcal{E})}}{\sqrt{2\mathcal{E}}}D_{-1-\rmi/(2 \mathcal{E})}\left((1+\rmi)\frac{q}{\sqrt{\mathcal{E}}}\right),\label{constfg}\\
g_c(q)=\frac{\rme^{-\pi/(8\mathcal{E})}}{\sqrt{2}}(1-\rmi)D_{-\rmi/(2\mathcal{E})}\left((1+\rmi)\frac{q}{\sqrt{\mathcal{E}}}\right).
\end{eqnarray}
\endnumparts
and $D_{\nu}(z)$ is the parabolic cylinder function.

\begin{figure}
\centering
\includegraphics[scale=0.9]{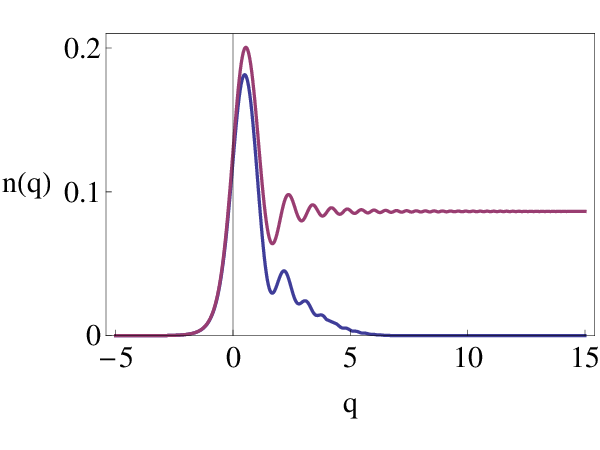}
\caption{The pair density $n(q)$ as a function of the dimensionless momentum $q$ for the constant (upper curve) and the exponential (lower curve) field for the dimensionless electric field strength $\mathcal{E}=1$. The value of the exponential field is taken at $\tau=0$ and for the value of $b=0.1$. Appropriate expressions are given in (\ref{n}) and (\ref{nqstat}).}\label{fig4}
\end{figure}

\begin{figure}
\centering
\includegraphics[scale=0.9]{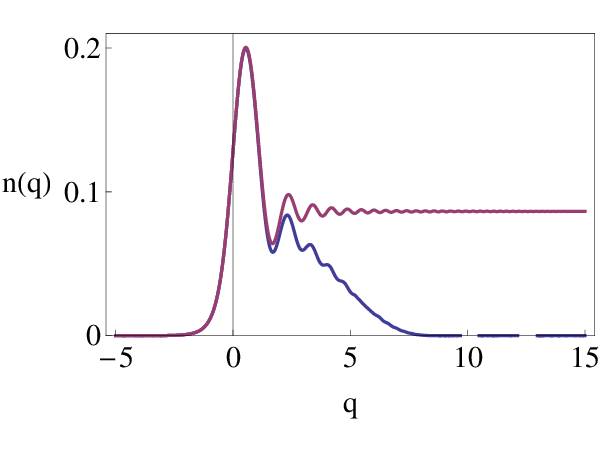}
\caption{The pair density $n(q)$ as a function of the dimensionless momentum $q$ for the constant (upper curve) and the Sauter (lower curve) field for the dimensionless electric field strength $\mathcal{E}=1$. The value of the Sauter field is taken at $\tau=0$ and for the value of $b=0.1$. Appropriate expressions for $n(q)$ are given in (\ref{nqstat}) and (\ref{n}) with $f(q,\tau)$ and $g(q,\tau)$ defined in (\ref{3sol}-{\it b}).}\label{fig5}
\end{figure}

In Fig.~\ref{fig4} we compare the pair density $n(q)$ for the constant and the exponential fields as functions of the dimensionless momentum $q$. The same comparison is shown in Fig.~\ref{fig5} for the constant field and the Sauter field. In both plots we took  $b=0.1$. The value of the time parameter was taken as $\tau=0$ because at this moment the field strength of the exponential or the Sauter field $\mathcal E$ is the same as that of the constant field, which was taken as $\mathcal{E}=1$. The pair density $n(q)$ for the exponential and the Sauter fields is given in (\ref{n}), while $f(q,\tau)$ and $g(q,\tau)$ for these fields are given respectively in (\ref{2sol}-{\it b}) and (\ref{3sol}-{\it b}).

Both plots clearly show the difference between the constant field and the exponential or the Sauter field. The pair density for all three fields is exactly the same for negative and small $q$, but for larger $q$ they are quite different. In the limit when $q\to \infty$ the pair density $n(q)$ for the constant field tends to a constant, which is nonanalytic in $\mathcal{E} = 0$. However, for the exponential and the Sauter fields the pair density tends fast to 0. Maximal rate of growth of the particles momenta caused by the exponential or the Sauter field equals $\mathcal{E}/b$. If we assume that pairs are created with $q \approx 0$, then the pair density in Figs.~\ref{fig4} and \ref{fig5} should be negligible for $q \leq 10$. Both plots satisfy this condition. It is also worth noting that for both the exponential and the Sauter field the oscillations of $n(q)$ for the constant field (\ref{nqstat}) are reproduced.

\begin{figure}
\centering
\includegraphics[scale=0.8]{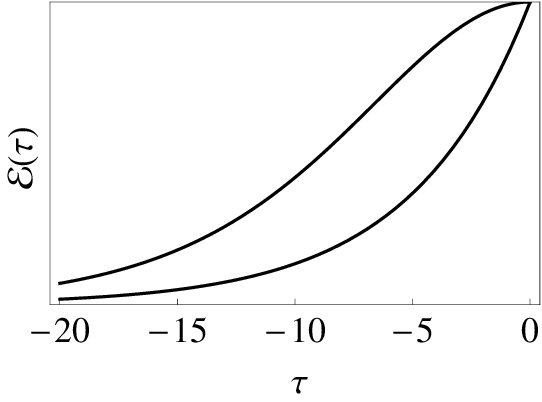}
\caption{The field strength $\mathcal{E}$ shown as a function of time. The lower curve corresponds to the exponential dependence and the upper curve to the square of the hyperbolic secans.}\label{fig6}
\end{figure}

\begin{figure}
\centering
\includegraphics[scale=0.8]{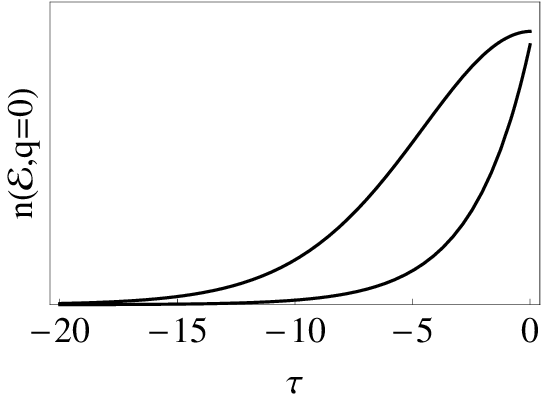}
\caption{The pair density $n(q,\tau)$ for the exponential field (lower curve) and the Sauter field (upper curve) plotted as a function of time. Since at $\tau=0$ both fields reach the same value, the number of pairs is almost the same.}\label{fig7}
\end{figure}

The two exactly solvable cases differ considerably but they share common features confirming the main message of our work. We have chosen the parameters in such a way that at $\tau=0$ the electric field in both cases attains the same value ${\mathcal{E}}$ and the rate of growth of the field in the remote past is $b$. The amplitude of the field is, however, four time larger for the Sauter field,
\begin{eqnarray}
\frac{\mathcal{E}}{\cosh^2(b\tau/2)}\approx 4\mathcal{E}\exp(b\tau).
\end{eqnarray}
In Fig.~\ref{fig6} we show the time dependence of the electric field in the two cases. There is a direct correspondence in the remote past between the functions that solve the two sets of equations. The confluent hypergeometric functions (\ref{2sol}-{\it b}) are related to the hypergeometric functions (\ref{3sol}-{\it b}) by the limiting procedure,
\begin{eqnarray}
_{1}F_1\left(\alpha,\gamma;z\right)
=\lim_{\beta \to \infty} {_{2}F_1}\left(\alpha,\beta,\gamma;z/\beta\right).
\end{eqnarray}
In our case evaluating the limit in (\ref{3sol}-{\it b}),
\begin{eqnarray}
\tau\to -\infty,\quad\mathcal{E}\to\infty,\quad8\mathcal{E}\rme^{b\tau}={\rm const}=b^2 s,
\end{eqnarray}
we obtain (\ref{2sol}-{\it b}).

\begin{figure}
\centering
\includegraphics[scale=0.8]{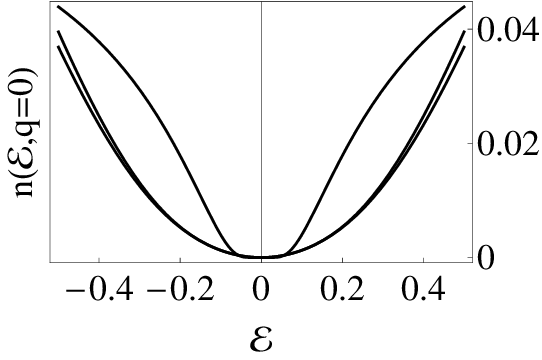}
\caption{The pair density $n(\mathcal{E})$ for the exponentially growing field at the time $\tau=0$ for $b=0.1$ and $q=0$ plotted as a function of the dimensionless field strength $\mathcal{E}$ for the exponential field (lower parabola) and for the Sauter field (upper parabola). For comparison we also show a typical nonanalytic behavior representing the Schwinger singularity.}\label{fig8}
\end{figure}

Since at $\tau=0$ the electric field attains in both cases the same value $\mathcal{E}$, we expect that the number of produced pairs will not differ significantly. Indeed, the results of the numerical calculations shown in Figs.~\ref{fig7} and \ref{fig8} fully confirm this expectation. The number of produced pairs is slightly higher for the Sauter field because, as shown in Fig.~\ref{fig6}, this field dominates over the exponentially growing field.

The analytic character of the number of pairs function is seen in Fig.~\ref{fig8} where we plotted the dependence on $\mathcal{E}$ in the neighborhood of zero. The parabolic character of both curves clearly shows that the expansion in powers of $\mathcal{E}$ begins in both cases with the quadratic term. The difference between the two cases diminishes with decreasing $b$.

\section{Conclusions}

We have shown with the help of exact analytic solutions that adiabatic switching of the electric field removes the essential singularity which is present in the Schwinger formula for pair creation. This observation opened the way to a perturbative treatment which gives the results coinciding with the power expansion of the exact solution. The method of spinorial decomposition that enabled us to find exact solutions is quite general and might be applicable to a class of similar problems.

\ack

All calculations and the figures we produced with the help of Mathematica \cite{math}. This research was partly supported by the grant from the Polish Ministry of Science and Higher Education for the years 2013--2016.

\appendix
\section*{Appendix}
\setcounter{section}{1}

The starting point is the expansion of the Dirac field operator (cf., for example, \cite{bd}) into creation and annihilation operators (c=1),
\begin{eqnarray}\label{dirf}
\fl \hat{\Psi}(\bm r,t)=\sum_s\int\frac{\rmd^3p}{(2\pi\hbar)^{3/2}}&\sqrt{\frac{m}{E_p}}\Big[u(\bm p,s)\hat{b}(\bm p,s,t)e^{\rmi\bm p\cdot\bm r/\hbar}+v(\bm p,r)\hat{d}^\dagger(\bm p,s,t)\rme^{-\rmi\bm p\cdot\bm r/\hbar}\Big].
\end{eqnarray}
We allowed for an arbitrary time dependence of the creation and annihilation operators, so that the validity of this formula is not restricted to free fields. Next, we invert the Fourier transform and use the properties of the spinors $u$ and $v$ to obtain
\begin{eqnarray}\label{dirf1}
\hat{b}(\bm p,s,t)=\sqrt{\frac{m}{E_p}}u^\dagger(\bm p,s)\!\cdot\!\int\frac{\rmd^3r}{(2\pi\hbar)^{3/2}}\rme^{-\rmi\bm p\cdot\bm r/\hbar}\hat{\Psi}({\bm r},t),
\end{eqnarray}
With the use of the completeness relation for the spinors \cite{bd},
\begin{eqnarray}\label{crel}
\sum_s u_\gamma(\bm p,s)u_\delta^\dagger(\bm p,s)=\frac{\left(\rho_3 m+{\rho_1\bm\sigma}\!\cdot\!{\bm p}+E_p\right)_{\gamma\delta}}{2m},
\end{eqnarray}
expressed in the original Dirac notation, we obtain the following formula for the number of electrons with momentum $\bm p$:
\begin{eqnarray}\label{nofel0}
\fl \hat{N}(\bm p,t)&=\sum_s\!\hat{b}^\dagger(\bm p,s,t)\hat{b}(\bm p,s,t)\nonumber\\
\fl &=\int\!\frac{\rmd^3r}{(2\pi\hbar)^3}\int\!\rmd^3\eta \rme^{-\rmi\bm p\cdot\bm\eta/\hbar}\hat{\Psi}^\dagger(\bm r-\bm\eta/2,t)\frac{\rho_3 m+\rho_1{\bm\sigma}\!\cdot\!{\bm p}+E_p}{2E_p}\hat{\Psi}(\bm r+{\bm\eta}/2,t).
\end{eqnarray}
Now, we replace the product of field operators by the commutator and anticommutator and use the canonical equal-time anticommutation relations
\begin{eqnarray}\label{anti}
\fl \hat{\Psi}_\beta^\dagger(\bm r-\bm\eta/2)\hat{\Psi}_\alpha(\bm r+\bm\eta/2)=\frac{1}{2}\delta(\bm\eta)\delta_{\alpha\beta}-\frac{1}{2}\left[\hat{\Psi}_\alpha(\bm r+\bm\eta/2),\hat{\Psi}_\beta^\dagger(\bm r-\bm\eta/2)\right].
\end{eqnarray}
We have not assumed that the field operators obey the free Dirac equation since we allowed in Eq.~(\ref{dirf}) for an arbitrary dependence of the creation and annihilation operators on $t$. However, in the presence of an external field we must introduce modifications required by gauge invariance. Following the procedure that led to the formula (\ref{dhw1}) we replace the shift operators in Eq.~(\ref{nofel0}) by their gauge invariant counterparts (\ref{rel}) and extract the phase factor $e^{-i\varphi}$. The gauge invariant number operator $\hat{N}_g$ obtained in this way is
\begin{eqnarray}\label{nofel}
\fl & \hat{N}_g(\bm p,t)=\sum_s\!\hat{b}^\dagger(\bm p,s,t)\hat{b}(\bm p,s,t)\nonumber\\
\fl &=\int\!\frac{\rmd^3r}{(2\pi\hbar)^3}\int\!\rmd^3\eta \rme^{-\rmi \left ( \bm p\cdot\bm\eta/\hbar + \varphi\right )}\hat{\Psi}^\dagger(\bm r-\bm\eta/2,t)\frac{\rho_3 m+\rho_1{\bm\sigma}\!\cdot\!{\bm p}+E_p}{2E_p}\hat{\Psi}(\bm r+{\bm\eta}/2,t).
\end{eqnarray}
The expectation value of ${\hat N}_g(\bm p,t)$ can then be expressed in terms of the DHW function. The quantity of interest is the pair density in phase space $n({\bm r},{\bm p},t)$ which in the present case is equal to the density of electrons. Therefore, in the formula (\ref{nofel}) we may drop the integration over $\bm r$ and the normalization of the phase-space volume element $(2\pi\hbar)^3$ to obtain the density of pairs per unit phase-space volume in the form
\begin{eqnarray}\label{nofel1}
n({\bm r},{\bm p},t)
=1+\frac{{\rm Tr}\left\{(\rho_3 m+\rho_1{\bm\sigma}\!\cdot\!{\bm p}+E_p){\bm W}({\bm r},{\bm p},t)\right\}}{2E_p}.
\end{eqnarray}
Due to the orthogonality of the Dirac matrices under the trace only the components $f_3,\bm g_1$, and $f_0$ may contribute and we obtain the final formula
\begin{eqnarray}\label{nofel2}
&n({\bm r},{\bm p},t)=1+\frac{m f_3({\bm r},{\bm p},t)+{\bm p}\!\cdot\!{\bm g}_1({\bm r},{\bm p},t)+E_pf_0({\bm r},{\bm p},t)}{2E_p}.\end{eqnarray}
The number of antiparticles calculated in the same manner is
\begin{eqnarray}\label{nofp2}
&{\bar n}({\bm r},{\bm p},t)=1+\frac{m f_3({\bm r},{\bm p},t)+{\bm p}\!\cdot\!{\bm g}_1({\bm r},{\bm p},t)-E_pf_0({\bm r},{\bm p},t)}{2E_p}.\end{eqnarray}
The difference between the two expressions is the charge density
\begin{eqnarray}\label{diff}
n({\bm r},{\bm p},t)-{\bar n}({\bm r},{\bm p},t)=f_0({\bm r},{\bm p},t),
\end{eqnarray}
as was to be expected.


\begin{thebibliography}{27}
\bibitem{err1} There is an editorial error in the corresponding equation in \cite{bbr}. The first ${\mathcal X}$ in (17) and also in the line below should be replaced by ${\mathcal H}$.
\bibitem{err2} There was a sign error in the corresponding matrix in \cite{bbr} in the first term in the upper row.
\bibitem{err3} This number was mistakenly reduced by a factor of 2 in \cite{bbr}.
\end{thebibliography}
\end{document}